\documentclass[12pt,aps,prb,preprint]{revtex4}   

\usepackage{amsmath}    
\usepackage{graphicx}   

\begin{document}

\title{Student experiences of virtual reality - \\a case study in learning special relativity}
\author{Dominic McGrath}
 \affiliation{Teaching and Educational Development Institute, The University of Queensland }
\author{Margaret Wegener, Timothy J. McIntyre}
\affiliation{School of Mathematics and Physics, Faculty of Science, The University of Queensland}
\author{Craig Savage, Michael Williamson }
\affiliation{Centre for Learning and Teaching in the Physical Sciences, The Australian National University}
\date{\today}

\begin{abstract}
We present a study of student learning through the use of virtual reality. A software package is used to introduce concepts of special relativity to students in a game-like environment where users experience the effects of travelling at near light speeds. From this new perspective, space and time are significantly different to that experienced in everyday life. The study explores how students have worked with this environment and how these students have used this experience in their study of special relativity.  A mixed method approach has been taken to evaluate the outcomes of separate implementations of the package at two universities. Students found the simulation to be a positive learning experience and described the subject area as being less abstract after its use. Also, students were more capable of correctly answering concept questions relating to special relativity, and a small but measurable improvement was observed in the final exam.
\end{abstract}

\maketitle

\section{Introduction}
Abstract physics is concerned with the development and analysis of conceptual models of physics, exploring the consequences of theory.  The aims of any course introducing an abstract physics topic to students include  exciting the students about the subject area, providing a rationale for abstract models, and developing abstract thinking abilities.  However many students undertake a surface learning approach\cite{Marton1976}, focusing on developing the ability to manipulate formulas in place of developing understanding of abstract principles.  ``I guess it is all mathematical formulas \ldots and we`re just like: OK, I don`t understand just sub in the things and we`ll get the answers right'' (Student focus group, 2008).  While this approach may enable students to achieve satisfactory results in an assessment task, these students have learnt little to support future study. 

A widely-used introduction to abstract physics is the study of special relativity; it is fundamentally and mathematically simple, popular and provides access to a fundamentally different understanding of time and space.  It is acknowledged, though, that many students fail to develop understanding of fundamental concepts in special relativity even after advanced instruction\cite{Scherr2001a}.  The problem is easily identified: ``It`s very bizarre and goes against what you know from real life, so it is very hard to grasp'' (Student focus group, 2008).  Understanding relativity requires one to accept that there is less that is absolute than was once believed and to accept a model of time and space that is strange and unfamiliar\cite{Mermin2005}. As such, modifying everyday concepts of motion, time and space to develop accurate constructs of the theory of special relativity is extraordinarily difficult\cite{Scherr2001b, Scherr2002}.
  
Previously published work describes common conceptual difficulties and misunderstandings, and activities to address these issues.  Mermin\cite{Mermin2005} and Scherr\cite{Scherr2001a} describe courses developed from research and experience; these courses support students` learning and avoid common pitfalls.  Ideas and recommendations for how and what special relativity should be taught in introductory courses have been featured in this journal\cite{Mallinckrodt1993, Mermin1994, Greenwood1982, Gjurchinovski2006}.

One avenue for student learning is laboratory work. However, experiments that support the study of special relativity are limited because the effects of special relativity only become significant at near light speeds.  Experiments have been developed to verify length contraction and time dilation by examining particles moving at near light speed\cite{Coan2006, Easwar1991, Lund2008}.  These experiments provide evidence to aid students` acceptance of special relativity but they use a heavily guided approach to ensure the collection of appropriate data for verification.   This approach is limited in its opportunities for student-led exploration of the concepts of special relativity.    

A viable alternative is the development of visualisations and simulations to aid conceptual understanding of special relativity through imagery and virtual experiences.  Gamow`s stories of Mr Tompkins in Wonderland\cite{Gamow1965} provide vivid descriptions of a relativistic world. More recently, computer-generated images and video that provide accurate imagery of motion at near light speed have been developed, for example, the work of Wieskopf et al\cite{Weiskopf2005}. Taylor\cite{Taylor1988} describes an educational implementation of several special relativity simulations, including a wire-frame three-dimensional simulation identifying visual effects of special relativity, and a model of a two-dimensional world featuring clocks.  Taylor`s two-dimensional simulation featured associated space-time diagrams, enabling exploration and connection between multiple representations of space and time.  Belloniet al\cite{Belloni2004} developed physlet simulations based on Scherr`s activities.  Carr et al\cite{Carr2007} have worked on developing serious games employing the effects of special relativity.  Recent technological advances enabled Savage et al\cite{Savage2007} to develop a virtual reality simulation, Real Time Relativity (RTR), modelling the visual, spatial and temporal effects of special relativity.  This software was implemented on easily available technology (personal computer with programmable graphics card) to provide the opportunity for new approaches to learning special relativity. 

This paper describes the implementation and evaluation of RTR and associated learning activities as performed iteratively over four semesters. The study builds on an earlier smaller scale study\cite{Savage2007} by accessing an expanded student cohort from two research-intensive institutions. The Australian National University (ANU) attracts students from around the country while students of The University of Queensland (UQ) tend to originate from the state of Queensland and from Asia. At these universities RTR has been embedded in first year physics courses where special relativity and quantum mechanics serve as an introduction to abstract physics. An important aspect of this approach is enabling students to recognise the value of abstract physics and develop a desire as well as a basis for further physics study.  

\section{Implementation}
The study began by using the existing version of RTR to form the basis of a laboratory activity at both participating institutions. Laboratory classes were chosen as they are the traditional setting for interaction and experimentation in physics courses. Laboratory experiences aim to provide students with concrete meaning, experience and representation to underpin theoretical understandings. Laboratory sessions enable groups of students to actively and collaboratively explore phenomenon to develop, test and utilise theory. Also, from a pragmatic perspective, the computers required to run RTR were most readily available in the teaching laboratories. 

\begin{figure}[ht]
\begin{center}
\includegraphics{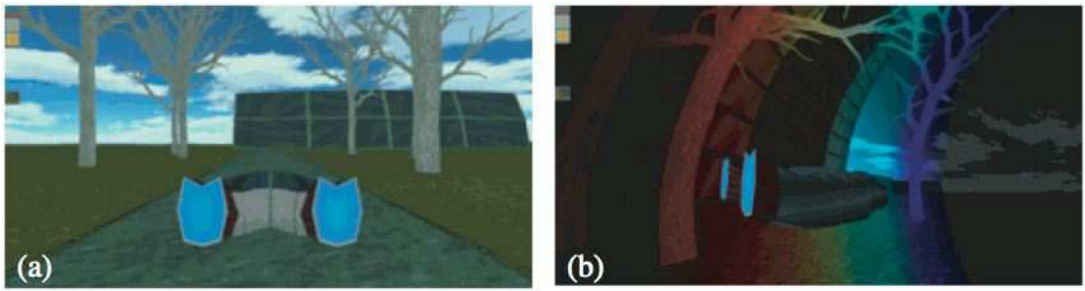}
\caption{\label{fig:1}Sample screen shots from the RTR software. (a) Cityscape scenario at zero relative velocity; (b) Moving through the cityscape at 0.866 times the speed of light.}
\end{center}
\end{figure}

The RTR software simulates the visual effects that become apparent when travelling at near light speed.  Users participate in a game-like world, with control over their motion and direction of view. Scenarios include environments consisting of clocks, planets and abstract shapes.  Some effects can be disabled in order to focus on particular concepts and also to allow the user to become accustomed to navigation controls and the scenarios; otherwise the visual experience of the user is extremely accurate.  The previous study had identified some difficulties with the navigational aspects of the software and this was addressed during the early stages of this implementation\cite{Savage2007}. Updated scenarios were also made available to the students. Sample screen images are shown in Figure \ref{fig:1}.

RTR was used by students in three-hour laboratory sessions. These classes contained small groups of students working in pairs. At ANU, students completed the experiment in parallel with a lecture series on special relativity. At UQ, the laboratory was part of a series of experiments that students completed on a rotational basis throughout the semester. Hence students may have attempted the experiment before, during or after the lecture and tutorial series on special relativity. Students at each institution were expected to prepare for the laboratory learning activity by reviewing materials available in textbooks and online in video and text formats, and to complete a series of preliminary questions.  On arrival, the responses were checked by tutors, who were also available to provide guidance throughout the session. 

The learning activities associated with RTR moved students through various stages, to manage the cognitive load and encourage exploration.  Initially students were encouraged to familiarise themselves with the interface and environment of RTR - exploring the various scenarios and controls available. Secondly, students were required to use RTR to connect selected theories and phenomena of special relativity and of the finite speed of light to their observations.  In this stage students` concepts were challenged and refined through comparison with observations and experiments in RTR.  Finally students had to apply their understanding to develop and carry out an experiment to verify the mathematical formulation of time dilation, an effect of special relativity. Throughout these learning activities students worked collaboratively, negotiating meaning and practice.  Working in an essentially visual desktop virtual reality, students were required to keep a log book record of their work in a visual manner, demonstrating the connections they made between abstract theory and visual effects. The log books were collected at the end of the session and assessed against the students` ability to demonstrate their understanding of the topics. 

Data about the student learning experiences was collected throughout 2008 and 2009, in Semester II at ANU and in Semester I and II at UQ. The study over this time involved more than 300 students (out of the 420 students enrolled in the relevant courses).  Students participating in this research were from a variety of disciplines, mainly  Engineering and Science, studying physics at first year university level.  For some students the course undertaken was required within an Engineering program.

Data collection targeted user confidence, self-efficacy and attitudes to physics, special relativity and technology, as well as level of performance on assessment tasks, and how students were learning. A variety of methods were used. Before and after completing their laboratory session, participants responded to closed and open survey questions, completed confidence logs\cite{Draper1996} and were administered concept-based multi-choice tests. In the concept tests, to avoid repetition effects and bias, sets of questions randomly assigned as pre-laboratory and post-laboratory were used. Survey questions probed beliefs of space and time which should be challenged by the study of special relativity. Randomly selected student groups were observed working in the laboratory, and informally interviewed when clarification was required.  Student focus groups reviewed the role of the laboraory activity in their learning, and preliminary survey findings.  Students were administered an online survey at the completion of their series of special relativity lectures and tutorials when approximately half the class had used RTR. Randomly-selected student laboratory work was reviewed. Performance on the final examination was compared for students who had completed RTR activities and those who had not. 

\section{Results}

We present the results of the study below, divided into sections on students` perceptions of the learning activities, learning outcomes based on concept tests and exam results, and the nature of the learning process.

\subsection{Student Attitudes}

A pre-laboratory survey probed students` attitudes and views of themselves in relation to their ability in understanding special relativity and their interests in using simulations as a learning tool. The results are shown in Table \ref{tab:1}, divided into responses from ANU and UQ cohorts. As expected, students indicated that they perceived special relativity as being a more abstract area of physics based on their previous experiences. There was a fairly neutral response to students` beliefs about their understanding of special relativity. UQ students rated slightly lower in this question, which may have been influenced by the rotational nature of the laboratory experiments in which some students attempted the experiment before the lecture series. UQ students were also more likely to believe that they could mechanically apply the equations of special relativity without having a good understanding of the associated physics.

\begin{table}[ht]
\begin{center}
\begin{tabular}{| p{13cm} | c  c | c  c |}
\hline
Pre-laboratory Question & \multicolumn{2}{|c|}{ANU} & \multicolumn{2}{|c|}{UQ} \\
& Mean & SD & Mean & SD \\
\hline
Special relativity is more abstract than other areas of physics. & 3.6 & 1.1 & 3.9 & 0.9 \\
I have a good understanding of special relativity. & 3.2 & 0.8 & 2.9 & 0.9 \\
I can use the formulae for special relativity but do not understand why they work. & 2.6 & 0.8 & 3.1 & 1.0 \\
I enjoy trying new things on a computer. & 4.0 & 0.8 & 3.9 & 0.9 \\
I find simulations are an effective way to learn. & 4.1 & 0.8 & 4.0 & 0.8 \\
\hline
\end{tabular}
\caption{\label{tab:1}Means and Standard Deviations for Pre-Laboratory Survey Results. Responses scored 1: Strongly disagree, 2: Disagree, 3: Neutral, 4: Agree and 5: Strongly agree. SD the standard deviation. Sample sizes were 31 at ANU and 146 at UQ. A 95\% confidence interval for the error in the means is less than 0.3 for all ANU values and 0.2 for all UQ values.}
\end{center}
\end{table}

Students were queried about their interests in using computers for learning and, in particular, their desire to use computer simulations to illustrate concepts in physics. The positive outcomes indicated that both institutions could consider using more simulations in their undergraduate courses.

\begin{table}[ht]
\begin{center}
\begin{tabular}{| p{13cm} | c  c | c  c |}
\hline
Post-laboratory Question & \multicolumn{2}{|c|}{ANU} & \multicolumn{2}{|c|}{UQ} \\
& Mean & SD & Mean & SD \\
\hline
I would like to learn more about special relativity. & 4.3 & 0.7 & 3.7 & 1.0 \\
I would be interested in using the Real Time Relativity software in my own time. & 3.6 & 0.7 & 3.0 & 1.1 \\
In other experiments it was easier to connect the theory to what I observed. & 2.9 & 1.1 & 3.1 & 1.1 \\
Using a relativity simulation is more fun than the other experiments. & 3.7 & 0.8 & 3.5 & 1.0 \\
I learnt more from this experiment than most others. & 3.3 & 0.9 & 3.4  & 1.0 \\
I would like to use more simulations in my studies. & 3.7  & 0.7 & 3.6  & 1.0  \\
I found this to be an interesting experiment.  & 4.0  & 0.6 & 3.9  & 0.8  \\
\hline
\end{tabular}
\caption{\label{tab:2}Means and Standard Deviations for Post-Activity Survey Results. Responses scored 1: Strongly disagree, 2: Disagree, 3: Neutral, 4: Agree and 5: Strongly agree. SD the standard deviation. Sample sizes were 31 at ANU and 175 at UQ. A 95\% confidence interval for the error in the means is less than 0.3 for all ANU values and 0.2 for all UQ values. }
\end{center}
\end{table}

A post-laboratory survey was completed by each student immediately after the learning activity. The results of this survey are shown in Table \ref{tab:2}. Students left the laboratory clearly interested in special relativity and with a desire to learn more about the topic. This was consistent with the earlier findings\cite{Savage2007}. Students also indicated that the simulation was more fun than other experiments; this might relate to the game-like nature of the simulation. ANU students indicated an interest in making further use of the RTR software in their own time; this motivation was not nearly as strong amongst the UQ students. The post-laboratory survey included a final response section for open comments. Although this was utilised by only about 15\% of respondents, of these, considerable numbers identified the activity as fun (20\% ANU, 48\% UQ) and helpful for understanding and learning (60\% ANU, 14\% UQ). 

Students were asked to complete Confidence Logs immediately before and after the laboratory activities. Students rated their own confidence with regards to various aspects of special relativity on a scale ranging from no confidence to very confident\cite{Draper1996}. Students` confidence ratings immediately prior to and immediately after completion of RTR laboratory activities were compared (using a paired  t-test to test the hypothesis of no improvement in confidence for each aspect). The results, shown in Table \ref{tab:3}, indicate that the groups of students at each university increased in confidence in all the aspects of special relativity that were measured. While indications of improved student confidence do not directly imply improvements in learning, they do indicate a change in students` perceptions, understandings and affective connection to a topic, which can flow on to improvements in learning.  A notable outlier in confidence development for UQ students was length contraction, an effect that is traditionally demonstrated in one or two dimensions and is quantified by a relatively simple mathematical formula.  RTR demonstrates length contraction in three dimensions, with the added complexities of other visual effects, challenging students to make complex connections between length contraction and other effects of special relativity. 

\begin{table}[ht]
\begin{center}
\begin{tabular}{| p{13cm} | c  | c  |}
\hline
Task & ANU & UQ \\
\hline
Explain the theory of special relativity to someone who isn`t studying physics & 1.0\% & $<$0.1\% \\
Apply aspects of the theory of special relativity to solve problems & $<$0.1\% & $<$0.1\% \\
Calculate the length contraction of a moving object given a relative velocity & 0.9\% & $<$3\% \\
Predict the change in colour of an object moving at near light speed & 0.1\% & $<$0.1\% \\
Describe the observed changes in shape of an object moving at high speed & $<$0.1\% & $<$0.1\% \\
\hline
\end{tabular}
\caption{\label{tab:3}Students` self-assessment of confidence level for specific tasks. The numbers show the probability of no improvement in confidence (measured by paired t-test) after completing the RTR exercise. Sample sizes were n=148 at ANU and 30 at UQ.}
\end{center}
\end{table}

\begin{figure}[ht]
\begin{center}
\includegraphics[width=0.8\textwidth]{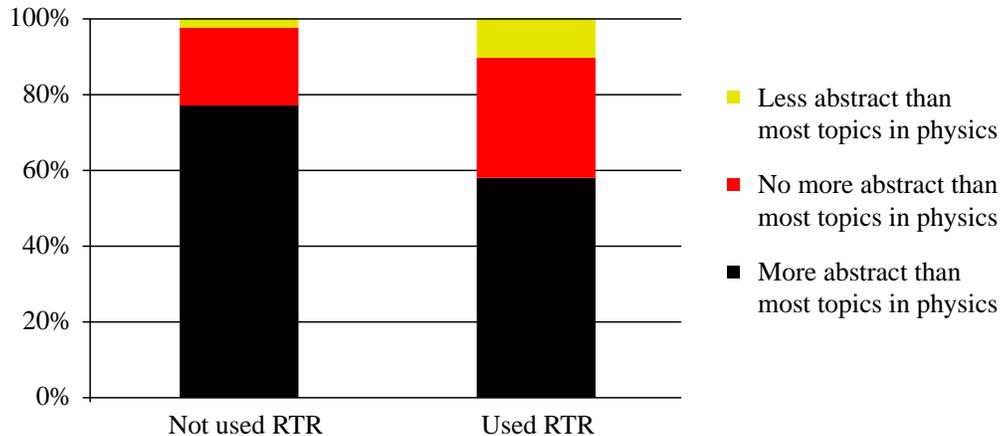}
\caption{\label{fig:2}Results from an on-line survey of students on the abstractness of special relativity. (ANU Semester II, 2008, n = 54) }
\end{center}
\end{figure}

One goal of the study was to provide a learning environment which presents special relativity in a less abstract way. Studies indicate that students perceive a topic area as abstract when it is presented as a mathematical model without meaning, application or connection to the physical world.  RTR provides a visual model to accompany the mathematical models of special relativity. The connection of special relativity to the normal human scale is made explicit as students experience relativistic effects at virtual speeds approaching that of light.  Through these experiences students identify special relativity as less abstract: ``It (special relativity) seemed more abstract without the prac'' (student Focus Group, 2008). The online survey also provides evidence of this change - see Figure \ref{fig:2}. Students who had used RTR, compared with those who hadn`t, were less likely to identify special relativity as being more abstract than most topics in physics; some even perceived it as less abstract than most physics topics (online survey).   

\subsection{Learning Outcomes}

Whilst students` perceptions of their level of learning are important in judging the effectiveness of a teaching innovation, it is (at least) equally important to examine the learning outcomes based on independent testing. To achieve this, students were presented with a short concept test before and after the RTR activity aimed at exploring their level of understanding of topics related to special relativity. The tests consisted of a set of belief statements requiring students to express their level of agreement or disagreement. The results of selected questions from the test are shown in Figure 3. The questions are divided into those where the statements are true (upper half) and those that are false (lower half).

\begin{figure}[ht]
\begin{center}
\includegraphics[width=\textwidth]{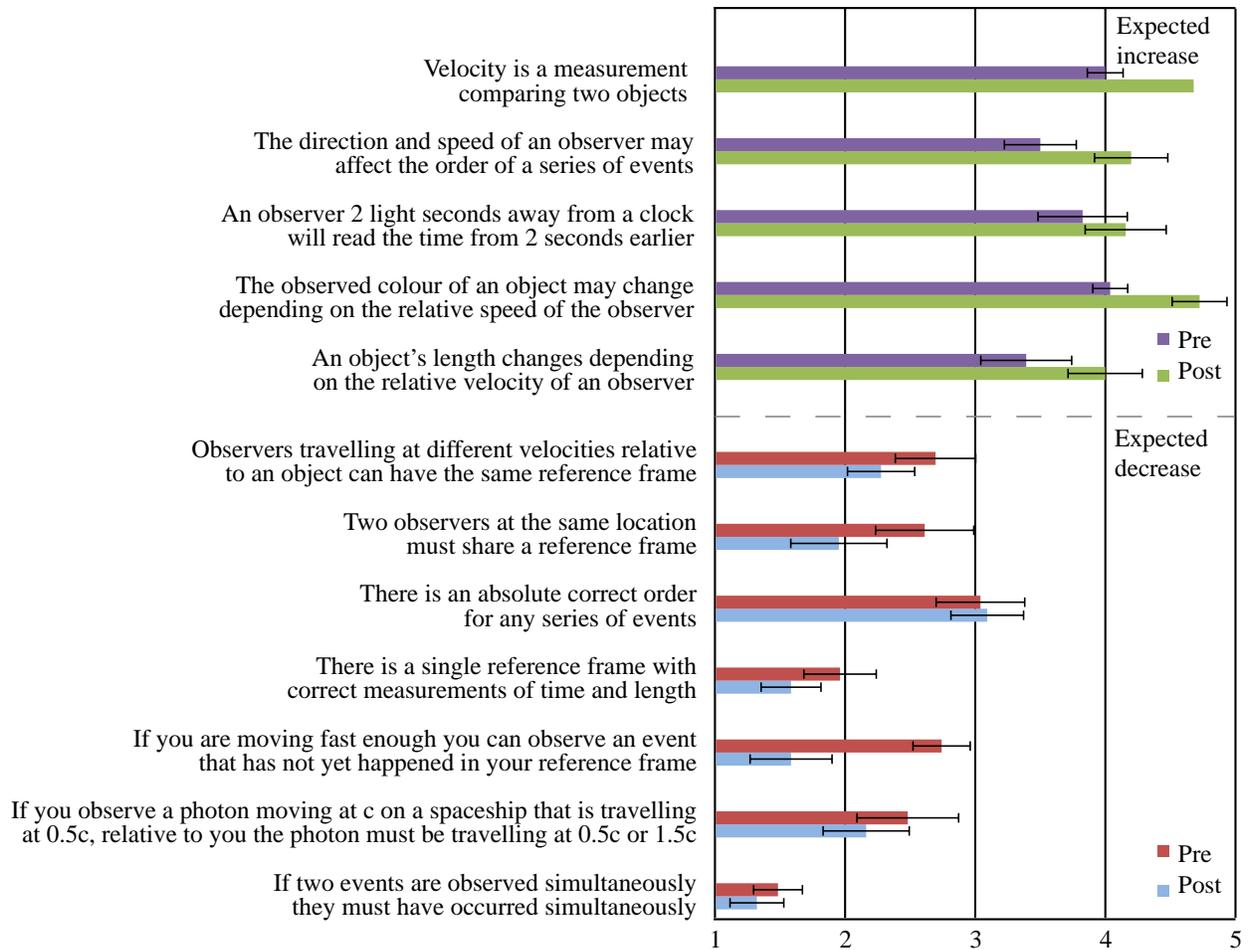}
\caption{\label{fig:3}Average of student responses to statements before and after the RTR activity. Responses scored 1: Strongly disagree, 2: Disagree, 3: Neutral, 4: Agree and 5: Strongly agree. (UQ Semester I, 2009, n=28) }
\end{center}
\end{figure}

Performance on these concept questions generally improved across a broad range of topics, showing an overall trend towards a deeper understanding of special relativity. In particular, questions concerning time dilation and simultaneity showed significant increased understanding (p$<$0.05). The improvement in time dilation was expected as time dilation is a specific focus for verification in the laboratory activities. The improvements in regard to simultaneity were unexpected as simultaneity was not an explicit focus for activities undertaken. Not all concept questions asked resulted in useful outcomes. In some cases, concepts were well understood prior to the laboratory activity, resulting in only an insignificant change in understanding. (These have not been included here.) In other cases, responses on less well understood topics showed no significant improvement - for example, the question on the existence of a correct order of events shown in Figure \ref{fig:3}.  The response rate of just under 60\% with a small cohort provided results that are statistically significant to within 5\% probability. Further research with a larger cohort and smaller question base may provide more conclusive data to substantiate the trends observed here. 

Student learning reflected the visual nature of RTR. This was indicated by feedback during focus groups held at the end of semester and in open-ended survey questions. Students reported gaining ``an ability to visualise relativistic effects which make it easy to apply theory '' (Student focus group, 2008), that ``it was much easier to learn the concepts (ALL concepts) of relativity when it is seen visually'' (open survey response), and ``it helped a lot with understanding because you could visualise something, that you have no experience of visualising in real life'' (open survey response).   These responses were reinforced when students explained how they approached problems in terms of visual models and examples from their experiences in RTR.  ``When we did the lab in here it reinforced all the ideas and also made it clearer -- Oh this is what happens you can visualise it `cause it is not something you can see every day'' (Student focus group, 2008).

Students at UQ were randomly assigned either to a group that completed the RTR experiment (n=134) or a control group who did not do the RTR experiment (n=51) due to the rotational structure of the laboratory sessions. For these two groups, responses on the final exam were compared. The final exam question on relativity was prepared and marked external to this study and was based on material presented in three one hour lectures and one tutorial session. Students were expected to conduct at least six hours of further independent study. It was found that students who completed the RTR experiment performed better on the special relativity question. More students in the RTR group performed at a higher standard than those who had not been assigned to undertake the RTR activities. While the effect size was small (Cohen`s d = 0.33), the effect was statistically significant (p$<$0.05 by unpaired t-test). A review of the exam papers did not reveal any significant difference in approach between these groups. Students who completed the RTR experiment were also found to have performed better on a question relating to quantum mechanics but no correlations were found with other topics. 

\subsection{Nature of Learning}

\begin{figure}[ht]
\begin{center}
\includegraphics[width=\textwidth]{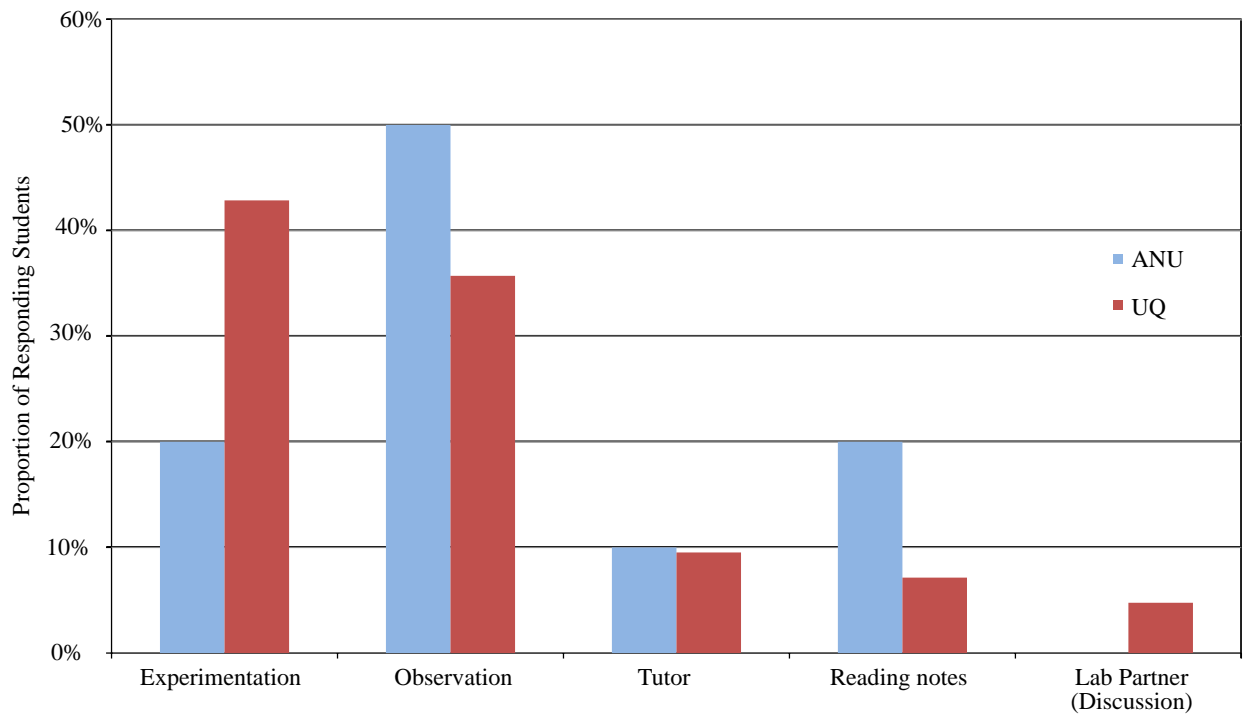}
\caption{\label{fig:4}Classified student open responses for how students learnt through the RTR activity. (ANU and UQ, n=52)}
\end{center}
\end{figure}

Responses to post-laboratory open questions about what and how students learned were classified using categories derived from the student responses. The classifications were reviewed and discussed by two members of the research team until consensus was reached. Students identified the most interesting aspect of the RTR experiment to be exploring an individual effect or a selected group of effects of special relativity (68\% ANU students, 80\% UQ students). Students learned about these effects through ``doing'' or travelling in RTR (20\% ANU, 45\% UQ), ``observing'' including seeing or watching (50\% ANU, 36\% UQ), tutor-led discussion (10\% ANU, 10\% UQ) and reading supplementary material (20\% ANU, 7\% UQ), demonstrating a predominantly active approach to learning, as shown in Figure \ref{fig:4}.  At least 70\% of responses from students at both institutions could be classified as typical descriptions of students undertaking hands-on experimental work. Thus virtual reality can be seen to be providing learning opportunities equivalent to other laboratory activities. Students also identified value in the active, student-led approach. For example: ``The section where you had to design an experiment about time dilation was useful \ldots you had to design and you weren`t just told what to do'' (Student focus group, 2008). Students also explicitly identified discussions with their laboratory partners as a way in which they learned.

Observations of student interaction during the laboratory sessions yielded further insights into the learning process. Students developed deeper understanding by negotiating the theoretical justification for their observations and challenging inaccurate conceptions through debate and experimentation.  Most students` initial experience of increasing speed involved moments of  ``Hang on, why are we going backwards?'' (UQ student using RTR), as aberration had a greater effect on perspective than motion.  Through a process of testing travel at a constant speed, increasing and decreasing speeds, students developed a concept of aberration.  Students then had to explicitly connect their experiences to theory by describing how these effects matched the theory of aberration. This matching of observation to labelled effects involved discussion within student groups, testing of concepts and guidance from laboratory tutors.  Students commented that ``initially it was overwhelming -- but it gets in your head by the end'' (UQ student, unstructured interview).  As students moved through various tasks in the laboratory activity, this progression through initial confusion to development of models of understanding, refinement and connection to theory was repeated.  The Doppler Effect, which was initially disabled for the experiment, inspired comments like ``It`s pretty but I don`t know what it is'' (UQ student using RTR) when turned on.  With the support of written material, peers and tutors, these conversations quickly moved to more mature analysis. For example, Student 1: ``Why is there a section losing colour?''  Student 2: ``It`s because infrared and ultraviolet and other wavelengths are outside the visible spectrum.'' (Student conversation in UQ laboratory session). These conversations consistently incorporated appropriate terminology and demonstrated the development of appropriate models.   

Students who used RTR after lectures or tutorials on special relativity reported benefits from their activity.  Experiences in RTR challenged misconceptions students brought to the laboratory which were not explicitly targeted in the laboratory activity.  A comment: ``I thought things would look bigger because you get smaller'' (ANU student using RTR) indicated a misconception regarding reference frames that the student then re-developed with the support of the RTR simulation, learning resources and their peers.  Students identified value in the focus on connecting the visual model represented in RTR and theory; ``I was able to refine my knowledge about some phenomenon, and also finally start assigning the right names to things.''  (open survey response), ``the lecturer was explaining aberration with the rain analogy, which I kind of understood but when I used RTR I went OK so that is what it is'' (Student focus group, 2008).  ``It helps show that while special relativity may seem counter-intuitive it is because our intuition was not developed travelling at near the speed of light''  (Student focus group, 2008). 

Students who used RTR before their series of lectures and tutorials also identified some benefits.  For example: ``I have a friend who did it [the experiment] about halfway through special relativity lectures and I was talking to him about it at lunch today, and [he] went, `Ah, now that you mention it', because when he did this experiment, he suddenly understood -- and that makes sense because I can pinpoint the time we were sitting in a row and would often be a little confused as a group of friends and think: it is this, but not quite be sure, and then halfway through he suddenly became more confident \ldots and now we know he did the experiment, so that makes sense'' (Student focus group, 2008). 

\section{Conclusions}

Real Time Relativity enables students to encounter phenomena that are outside human experience. The simulation along with an accompanying instructional package has been successfully implemented and studied in an undergraduate laboratory setting at two tertiary institutions. Data was collected through pre- and post-laboratory surveys and concept tests, through focus groups, through observation of student interactions during laboratory sessions and through analysis of exam papers completed by the students. 

The study showed that students were able to develop visual models of the effects of special relativity through the use of the simulation. This model aided their understanding and enabled the students to see the topic as less abstract. Students enjoyed the learning experience and became more confident in their understanding of the topic. Post-laboratory evaluations showed that not only did students perceive that they had an improved understanding, they were also able to perform better in concept tests about special relativity. Preliminary results showed that completing the experiment aided in answering an exam question based on the lecture component of the course. 

In the implementation at the two universities there were some variations in the student types and in the presentation of other relevant learning activities. The fact that the package was robust enough to yield positive responses at each institution, overseen by different staff teams, gives confidence that the RTR simulation and teaching package would perform favourably in wider implementation.

The outcomes of this work show that virtual reality can aid in changing beliefs about reality and can assist in making traditionally difficult abstract topics more accessible. We are currently investigating other topics where a visual approach may aid in student learning, and have begun work on a simulation of concepts in quantum mechanics.

\begin{acknowledgments}
Support for this study has been provided by The Australian Learning and Teaching Council, an initiative of the Australian Government Department of Education, Science and Training. The views expressed in this paper do not necessarily reflect the views of The Australian Learning and Teaching Council. 

The Real Time Relativity simulator was originally developed at ANU by Lachlan McCalman, Anthony Searle and Craig Savage. 

Real Time Relativity is available free from http://realtimerelativity.org/ 
\end{acknowledgments}

\end{document}